\documentclass[prd,nofootinbib,showpacs,preprint,floatfix]{revtex4}
\usepackage{latexsym}
\usepackage{dcolumn}
\RequirePackage{graphicx}

\newcommand{\MET}{\slash\!\!\!\!E_T}
\newcommand{\DZero}{D$\slash\!\!\!0$}
\newcommand{\fbi}{fb$^{-1}$}
\newcommand{\Za}{\widetilde{\chi}_1^0}
\newcommand{\Zb}{\widetilde{\chi}_2^0}
\newcommand{\Zc}{\widetilde{\chi}_3^0}
\newcommand{\Wpa}{\widetilde{\chi}_1^+}

\newcommand{\Slp}{\widetilde{l}}

\begin{document}


\preprint{SMU-HEP-08-01, ANL-HEP-PR-08-21, NSF-KITP-08-56}

\title{Trilepton production at the CERN LHC: Standard model sources and beyond}

\author{Zack~Sullivan}
\affiliation{Southern Methodist University,
Dallas, Texas 75275-0175, USA}
\affiliation{High Energy Physics Division,
Argonne National Laboratory,
Argonne, Illinois 60439, USA}
\author{Edmond~L.~Berger}
\affiliation{High Energy Physics Division,
Argonne National Laboratory,
Argonne, Illinois 60439, USA}

\date{May 23, 2008}

\begin{abstract}
Events with three or more isolated leptons in the final state are known to be
signatures of new physics phenomena at high energy collider physics
facilities.  Standard model sources of isolated trilepton final states include
gauge boson pair production such as $WZ$ and $W\gamma^{*}$, and $t\bar t$
production.  We demonstrate that leptons from heavy flavor decays, such as $b
\rightarrow l X$ and $c \rightarrow l X$, provide sources of trileptons that
can be orders of magnitude larger after cuts than other standard model
backgrounds to new physics processes.  We explain the physical reason heavy
flavor backgrounds survive isolation cuts.  We propose new cuts to control the
backgrounds in the specific case of chargino plus neutralino pair production
in supersymmetric models.  After these cuts are imposed, we show that it
should be possible to find at least a $4\sigma$ excess for supersymmetry
parameter space point LM9 with 30 \fbi\ of integrated luminosity.
\end{abstract}


\maketitle

\section{Introduction}
\label{sec:introduction}

Events with three or more isolated leptons in the final state have attracted
significant attention as possible signatures of new physics phenomena produced
at high energy collider physics facilities such as the Fermilab Tevatron
\cite{CDFDZ} and the CERN Large Hadron Collider (LHC)
\cite{CMSTDR2,CMSpaper,ATL97,ATLANAL,Vandelli:2007zza}.  To the extent that
new physics at the scale of $100$~GeV or more is related to electroweak
symmetry breaking, new states predicted in beyond the standard model (BSM)
schemes typically couple to the $W^{\pm}$ or $Z^0$ gauge bosons which, in
turn, decay to leptons.

Supersymmetry (SUSY) is a well studied potential realization of physics beyond
the standard model~\cite{SUSYreview}.  Trilepton events have long been
mentioned as potential golden discovery modes for SUSY processes in which a
chargino $\widetilde{\chi}_i^+$ and a neutralino $\widetilde{\chi}_j^0$ are
jointly produced.  Isolated three lepton signatures arise from the decay modes
$\widetilde{\chi}_i^+ \to W^{*\,+} \Za \to \bar{l}\nu\Za$, along with
$\widetilde{\chi}_j^0 \to Z^*\Za \to l\bar{l}\Za$.  In this example, the $\Za$
is the lowest mass SUSY particle assumed to be effectively stable and
therefore a source of missing energy; $W^{*\,+}$ and $Z^*$ stand for on-shell
or virtual gauge bosons.  This SUSY example provides a relatively clean source
of isolated trilepton events~\cite{cleantrilept,backgrounds}.  Within SUSY,
one can also pair-produce heavy sparticles that carry the color quantum
number, such as gluinos and squarks.  The cascade decays of these states,
while parameter dependent, will generally result in trilepton final states
along with hadronic jets~\cite{leptplusjets}.

Other BSM constructs, such as Higgsless models, extra-dimensional models, and
models with extended gauge groups, predict states at high mass, such as
Kaluza-Klein towers and extra gauge bosons, that decay to the more familiar
$W$ and $Z$ electroweak gauge bosons.  In one example of a Higgsless model, a
massive $W'$ is predicted with dominant decays into pairs of gauge bosons,
such as $W^{\pm} Z$, which is a source of isolated three lepton final
states~\cite{He:2007ge,Wprime,KK}.  Our intent here is not to present a survey
of BSM sources of multilepton events.  We cite the cases mentioned as
indicative of a range of possibilities.

In this paper we are concerned with obtaining as good an estimate as possible
of the production of isolated three lepton final states that arise entirely
from sources within the standard model (SM) itself.  An obvious and important
example is the associated production of a pair of gauge bosons, such as $WZ$,
along with its generalizations $W\gamma^{*}$, where $\gamma^{*}$ is a virtual
photon that decays as $\gamma^{*}\rightarrow l\bar{l}$.  While the importance
of the $W\gamma^{*}$ process has been emphasized as an important background
before \cite{backgrounds,Campbell:1999ah}, we demonstrate that its
contribution significantly changes the expectations of the particular CMS and
ATLAS studies we examine.  Unlike $WZ$, the $W\gamma^{*}$ contribution cannot
be reduced by an antiselection of events in which the invariant mass of the
$l\bar{l}$ system is in the vicinity of the $Z^0$ peak.

In this work, we use MadEvent~\cite{Maltoni:2002qb} to compute the full matrix
elements for partonic subprocesses that result in a $lll\nu$ final state.  For
the $WZ/\gamma^{*}$ process, this method allows us to retain important angular
correlations among the final leptons.  Here $ Z/\gamma^{*}$ stands for the
full range of $l\bar{l}$ pairs resulting from a $Z$, a Drell-Yan virtual
photon, and photon-$Z$ interference.  Our estimates are larger for this
process than those one would obtain from a modeling of $WZ$ alone from within
PYTHIA~\cite{PYTHIA}, and we describe these differences.  Inclusion of the
$\gamma^{*}$ contribution accounts for part of these differences.  Since
$WZ/\gamma^{*}$ is an important standard model background in many searches for
new physics, we suggest that events be generated by a matrix element program
and be fed into the PYTHIA showering routines, rather than use of the $WZ$
routine built into PYTHIA.

A major new contribution in this paper is the demonstration that isolated
leptons from bottom and charm decays are a potent source of standard model
backgrounds to new physics signatures in the three lepton final state.  This
study is an extension of our prior investigation~\cite{Sullivan:2006hb} of SM
heavy flavor backgrounds to the isolated two-lepton final state that is
important in searches for the Higgs boson.  In this new paper we compute
contributions from a wide range of SM heavy flavor processes including $b
Z/{\gamma^{*}}$, $c Z/ \gamma^{*}$, $b \bar{b} Z/ \gamma^{*}$, $c \bar{c} Z/
\gamma^{*}$.  We also include contributions from $t \bar{t}$ production, and
from processes in which a $W$ is produced in association with one or more
heavy flavors such as $t W$, $b \bar{b} W$, $c \bar{c} W$.  In all these
cases, one or more of the final observed isolated leptons comes from a heavy
flavor decay.  The $b \bar{b} W$ and $c \bar{c} W$ contributions have not been
examined previously.

To examine signal discrimination (and to compute signal to background ratios),
we choose the specific case of chargino and neutralino pair production as the
signal process.  We focus on the SUSY parameter space points LM1, LM7, and LM9
considered by the CMS Collaboration~\cite{CMSTDR2,CMSpaper} and on the SU2
point studied by the ATLAS Collaboration~\cite{ATLANAL}.  These points are
expected to have favorable SUSY cross sections at the LHC.  The dominant
nature of some of the SM backgrounds motivates the investigation of new
selections (cuts) on the final state kinematic distributions that would be
effective in reducing the backgrounds.  One of these cuts involves selections
on the opening angles among the three charged leptons in the final state. The
cuts are defined and their effectiveness is discussed in Secs.\ \ref{sec:CMS}
and \ref{sec:ATLAS}.

We begin in Sec. II with an explanation of the source and magnitude of
isolated leptons from heavy flavor production and decay.  Issues important in
our simulation of final states are discussed in Sec. III.  We generate events
using MadEvent, pass them through a PYTHIA showering Monte Carlo code, and
finally through a heavily modified PGS detector simulation program.  Our
overall results and comparisons with studies done by the CMS and ATLAS groups
are presented in Secs.\ \ref{sec:CMS} and \ref{sec:ATLAS}.  Our conclusions
are summarized in Sec.\ \ref{sec:conclusions}.  We show that it should be
possible to find at least a $4\sigma$ excess in 30 \fbi\ for SUSY point LM9.

\section{Isolated leptons from bottom and charm decays}
\label{sec:physbd}

The use of leptons inside jets to tag bottom and charm jets has led to the
recognition that strong isolation criteria are required to distinguish leptons
from primary interactions from those caused by secondary decays.  In Ref.\
\cite{Sullivan:2006hb} we demonstrate that the rate for secondary leptons to
pass isolation cuts is surprisingly large --- about 1 part in 200.  In this
section we explain the reasons these leptons pass the isolation cuts at such a
large rate.

\begin{figure}[htb]
\centering
\includegraphics[width=3.25in]{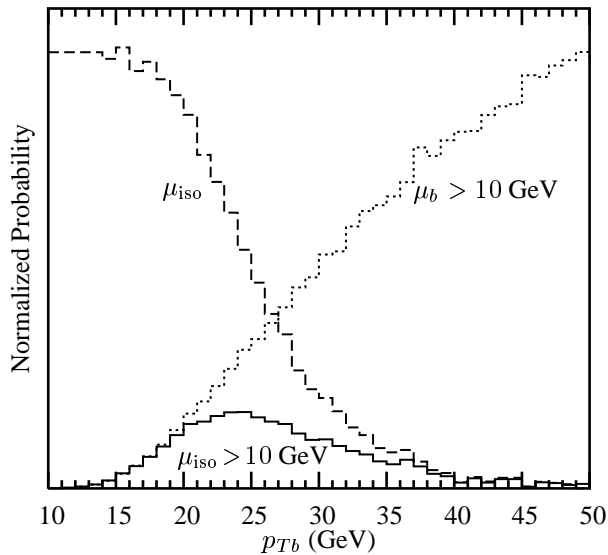}
\caption{Normalized probability for an existing $b$ quark to produce an
isolated muon with $p_{T\mu}>10$ GeV (solid) vs.\ the $b$ transverse momentum.
This curve is a multiplicative combination of the probability of producing a
muon with $p_{T\mu}>10$ GeV (dotted) and the probability the muon will be
isolated (dashed).  The $b$ production spectrum is not included.  
Muon isolation criteria are described in the Appendix.
\label{fig:pmuptb}}
\end{figure}

The simplest case to understand is the production of muons from $b$ quark
decay.  The branching fraction of various $B$ hadrons ($B_d$, $B_s$,
$\Lambda_b$, etc.) to muons is roughly 9--14\%.  Observation of these leptons
requires, however, that their charged tracks generally be above some
transverse momentum threshold (see the Appendix \ref{app:isolep} for details
of isolated lepton reconstruction).  In the dotted line of Fig.\
\ref{fig:pmuptb} we show the shape of the probability to produce a muon with
$p_{T\mu}>10$ GeV versus the transverse momentum of an initial $b$ quark.
This curve is determined predominantly from the $V-A$ matrix element decay of
the $B$ hadron into $Dl\nu$~\cite{PYTHIA} and kinematics imposed by the muon
momentum threshold.\footnote{There is a small dependence in the shape due to
the default choice of Peterson fragmentation for $b\to B$.  A slightly harder
$B$ spectrum is predicted by a nonperturbative fragmentation function
\protect\cite{Cacciari:2002pa}, which in turn predicts that the leading edge
of the dotted line in Fig.\ \protect\ref{fig:pmuptb} would shift 1--2 GeV
lower.  This implies softer $b$ quarks can produce 10 GeV muons, and hence
would slightly \textit{increase} the backgrounds compared to those calculated
here.}  The probability of producing a 10 GeV muon is small at low $b$
transverse momentum, and it grows toward the branching fraction limit as the
transverse energy of the $b$ moves farther above threshold.  This curve leads
to the false impression that low-$p_{T}$ $b$ quarks are unimportant.

There are two ways in which a muon can pass isolation cuts applied to detector
data.  The first is if the other decay products of the $B$ hadron are
physically separated in pseudorapidity $\eta$ and/or radial angle $\phi$.
This separation accounts for no more than 1/2 of the isolated rate.  One could
imagine rejecting secondary muons by increasing the area examined for
accompanying radiation, but this technique reduces acceptance of primary muons
and is of limited utility.  Of more importance is the isolation energy cut.
When muons come from low-$p_{T}$ $b$ quarks, they must have taken most of the
transverse momentum of the $B$ hadron.  This means there is not enough energy
left in the other $B$ decay products to fail the energy cuts for isolation.
The net isolated muon probability is the multiplication of the probability for
production and for passing isolation cuts.  This probability is shown as the
solid line of Fig.\ \ref{fig:pmuptb}, and it peaks fairly close to threshold.

There is little freedom to change the picture in Fig.\ \ref{fig:pmuptb}.  An
attempt could be made to lower the energy threshold beyond which events are
rejected, but this cut is already nearly optimized for the acceptance of real
muons.  Hence, one might expect to reduce the rate of muons from
higher-$p_{T}$ $b$ decays, but there will not be much gain.  One other handle
might be to look for secondary vertices, but our tests indicate that virtually
all muons that pass isolation point back at the primary vertex.  A preliminary
examination of CDF data \cite{privcom1} appears to confirm this finding.

\begin{figure}[htb]
\centering
\includegraphics[width=3.25in]{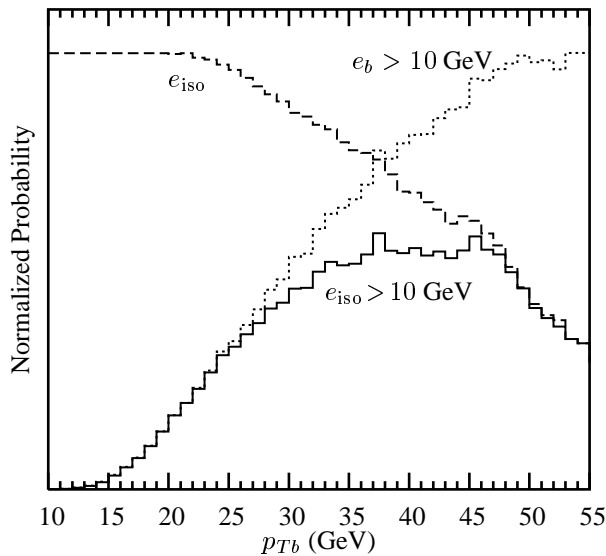}
\caption{Normalized probability for an existing $b$ quark to produce an
isolated electron with $p_{Te}>10$ GeV (solid) vs.\ the $b$ transverse
momentum.  This curve is a multiplicative combination of the probability of
producing an electron with $p_{Te}>10$ GeV (dotted) and the probability the
electron will be isolated (dashed).  The $b$ production spectrum is not
included.   Electron isolation criteria are described in the Appendix.  
\label{fig:peptb}}
\end{figure}

The physics behind finding an isolated electron from $b$ quark decay is
similar to that for finding a muon.  Figure \ref{fig:peptb} demonstrates the
same probability for production of an electron.  The probability of satisfying
the isolation cuts is flatter with respect to $p_{Tb}$ than for the muon,
because noise in the electromagnetic calorimeter requires less stringent cuts
to maintain electron acceptance. It is possible to achieve some purity with a
loss of acceptance, but we find little difference in the net rate of leptons
from heavy flavors between the ATLAS reconstruction algorithm that we
use~\cite{ATLLEPS} and less sophisticated algorithms.

\begin{figure}[htb]
\centering
\includegraphics[width=3.25in]{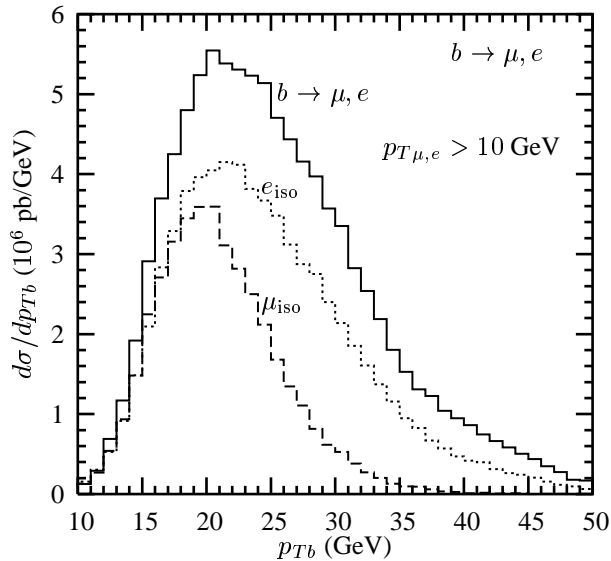}
\caption{Cross section for production of a muon or electron from $b\bar b$
production and decay (solid), an isolated muon (dashed), or an isolated
electron (dotted).
\label{fig:lsigptb}}
\end{figure}

While physical limitations of the experimental apparatus control the
acceptance of a lepton produced from a $b$ decay, the overall rate of
reconstructed leptons includes the effect of the $p_{Tb}$ production spectrum.
With the exception of top-quark decay, the $p_{Tb}$ spectrum tends to fall
steeply.  In Fig.\ \ref{fig:lsigptb} we display the net isolated lepton cross
section as a function of $p_{Tb}$ for $b\bar b$ production.  Despite the
larger acceptance for isolated muons around $p_{Tb}\sim 25$ GeV and isolated
electrons with $p_{Tb}\sim 30$--50 GeV, the peak production comes from $b$
quarks around 20 GeV.  The position of this peak has a profound effect on the
simulation of isolated leptons, because fully 1/2 of the reconstructed muon
rate comes from $b$ quarks with $p_{Tb} < 20$ GeV.  A natural tendency is to
ignore these lower energy $b$ quarks based on the low probability of lepton
production shown in the dotted lines of Figs.\ \ref{fig:pmuptb} and
\ref{fig:peptb}.  The complete picture, however, demonstrates that these
events are important and most difficult to reject.  The net effect is that
about 1\% of $b$ quarks produce a lepton ($\mu$ or $e$) that passes isolation
cuts.  This agrees with a \DZero\ study using their full detector simulation
that found $\sim0.5\%$ of $b$ hadrons produced isolated muons
\cite{Abazov:2004au}.

\begin{figure}[htb]
\centering
\includegraphics[width=3.25in]{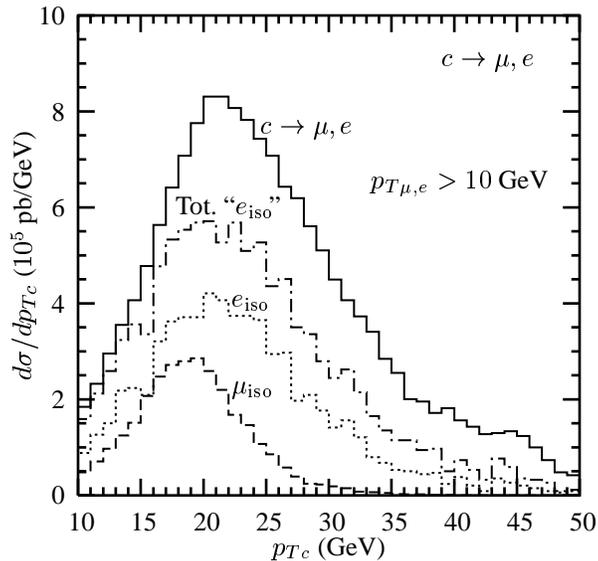}
\caption{Cross section for production of a muon or electron from $c\bar c$
production and decay (solid), an isolated muon (dashed), an electron which is
then isolated (dotted), or ``isolated electron'' including fakes (dash-dot).
\label{fig:lsigptc}}
\end{figure}

An examination of charm decay into isolated muons and electrons leads to
essentially identical conclusions as bottom decay with one minor addition.  In
Fig.\ \ref{fig:lsigptc} we show the cross section for production of a muon or
electron vs.\ $p_{Tc}$, and their isolated rates.  In practice, because of
``fakes'' (charged hadrons mistaken for electrons), the reconstructed rate for
electrons is larger than expected from the decay of $c$ to $e$ times an
isolation acceptance.  A large number of charged pions can be produced in
decays of $D$ mesons.  A reasonable fraction of these pions are reconstructed
as electrons, and so the net rate for ``isolated electrons'' shown as the
dot-dashed line is about 50\% larger than expected.  Electron reconstruction
algorithms are constructed in order to balance this contamination of charged
pions versus acceptance.  The goal across detectors is typically a fake rate
for pions from jets of $10^{-4}$ (compared to the rate of leptons we find of a
few times $10^{-3}$ from $b$ and $c$ decays) \cite{ATLLEPS, CMSTDR}.
Depending of the physics study, it may be worth including the decays from
heavy quarks as a part of the design of electron isolation algorithms.  For
the case of trileptons we examine here, the leptons from heavy quarks dominate
the backgrounds.

We conclude this section by noting that Figs.\ \ref{fig:lsigptb} and
\ref{fig:lsigptc} indicate the expected rate of isolated electrons from
processes containing $b$ and $c$ quarks can be significantly larger than the
rate for isolated muons.  Consequently, new physics processes with muons
should have a higher purity than those with electrons.  One general strategy
would be to look for signals with an excess that is more apparent in muon
channels than electron channels.  A second observation is that looser cuts
cause the typical transverse momentum of surviving isolated electrons to be
slightly higher than that for muons.  This recognition suggests that electron
events will be more suppressed than muon events by cuts on maximum lepton
energy, e.g., the $Z$-peak cut used below.  This lepton flavor dependence may
allow an additional handle on \textit{in situ} calibration of background
sources.

\section{Simulation}
\label{sec:sim}

The backgrounds we discuss tend to arise from the tails or end points of
physical processes and challenge detector capabilities.  To achieve a
believable and detailed simulation of reconstructed events, we follow the
methods developed and validated in Ref.\ \cite{Sullivan:2006hb}, with some
modifications for efficient production and analysis of the results.

We generate events with a customized\footnote{We use a new ``zooming''
procedure that improves phase space filling.}  version of MadEvent 3.0
\cite{Maltoni:2002qb} and run them through the PYTHIA 6.327 \cite{PYTHIA}
showering Monte Carlo.  Both programs use the CTEQ6L1 parton distribution
functions \cite{Pumplin:2002vw} evaluated via an efficient evolution code
\cite{Sullivan:2004aq}.  The showered events are fed through a version of the
PGS 3.2 \cite{Carena:2000yx} fast detector simulation, modified to match ATLAS
geometries, efficiencies, and detailed reconstruction procedures
\cite{ATLLEPS}.  We confirm that our results agree with full ATLAS detector
simulations at both the reconstructed object (leptons, jets, $\MET$) and
analysis levels to better than 10\% (and in some cases better than 1\%).

For this paper we use the ATLAS muon and electron reconstruction procedure
(described in the Appendix) and detector efficiencies to produce initial
candidate leptons for both CMS and ATLAS studies.  We then apply the CMS or
ATLAS geometric acceptance, transverse momentum threshold, and analysis cuts
in order to compare to their to their respective studies.  We reproduce the
results of the published CMS analysis to better than 1\% in most cases.  As we
expect from Sec.\ \ref{sec:physbd}, and as observed in Ref.\
\cite{Sullivan:2006hb}, we find that since the detectors have been optimized
for the same lepton acceptance and lepton fake rate from jets, the lepton rate
from heavy-flavor decays is insensitive to details of the detector
reconstruction.

In Sec.\ \ref{sec:physbd} we explain the physics reasons for simulating
low-$p_T$ $b$ and $c$ quarks.  In Ref.\ \cite{Sullivan:2006hb} we determined
that at least 1000 samples of phase space are required to reproduce the
interesting angular correlations in phase space for these types of events.  As
missing energy plays an important role in these signals, we cannot just
multiply the spectrum of produced quarks by an average lepton isolation
acceptance.  Instead, we iterate over each point in phase space between $10^4$
and $5\times 10^5$ times, ensuring we produce at least 10 events that pass a
minimal set of cuts.  This number is sufficient if there are two $b$ or $c$
quarks that are needed to produce leptons.

Simulating any cross section that requires 3 or more heavy quarks to produce
leptons ($bbbb$, $cccc$, $bbcc$) would require at least 1000 times the number
of events we simulate.  Hence, we can provide only an order-of-magnitude
estimate of the effect of these four heavy-flavor processes in Sec.\
\ref{sec:CMS}.

Several technical hurdles have to be overcome to achieve the statistically
significant and systematically controlled results shown in Secs.\
\ref{sec:CMS} and \ref{sec:ATLAS}.  We have already described our method of
iterating multiple times over the same phase space points to give showering an
opportunity to generate isolated leptons.  Even though the detector simulation
is technically a ``fast'' detector simulation, over 10 billion events were
required to produce sufficient statistics to describe the heavy flavor decays
to isolated leptons throughout their available phase space.  This computation
was carried out with 2 CPU years of time on the Argonne Laboratory Computing
Resource Center JAZZ cluster, a Pentium Xeon Linux cluster with 350 compute
nodes.  The large number of events raises a practical issue regarding the
total amount of data generated.  Because of the time involved in generation,
and our desire to analyze both angular correlations and different cuts for CMS
and ATLAS, we must store event information.  Standard PYTHIA, STDHEP, or
compressed data structures for this analysis would require more than 100
terabytes of storage, a simply impractical requirement.  Therefore, we write
only events that produce at least 3 isolated leptons (see the Appendix for the
definition of isolation), and store the four-vectors of the leptons, leading
jet (for vetoes), and missing transverse energy.  This is the minimal
necessary information to analyze the data presented below.

\section{Comparison to CMS}
\label{sec:CMS}

In their technical design report (TDR) \cite{CMSTDR2}, the CMS Collaboration
defines several several sets of minimal supergravity (mSUGRA) parameters that
have sensitivity to different aspects of SUSY.  The points LM1, LM7, and LM9
are the only subset of the 9 points examined that exhibit a large trilepton
signature from $\Zb\widetilde{\chi}_1^\pm$ decay.  These points assume
$A_0=0$, $\mu>0$, and ($m_0$, $m_{1/2}$, $\tan\beta$) are (60, 250, 10),
(3000, 230, 10), and (1450, 175, 50) for LM1, LM7, and LM9, respectively.  The
reach for discovery of $\Zb\widetilde{\chi}_1^\pm$ decays to trileptons is
summarized in the TDR, but we follow the more detailed account from Ref.\
\cite{CMSpaper} below.

The basic CMS analysis looks for 3 isolated leptons, with $p_{T\mu}>10$ GeV,
$p_{Te}>17$ GeV, and $|\eta_l|<2.4$ (see the Appendix for the method of
isolation).  Events that contain jets with $E_{Tj}>30$ GeV are vetoed in order
to reduce the contribution from processes such as $t \bar{t}$ production and
more complex SUSY processes involving cascade decays of massive SUSY states.
Since real $Z$ decays dominate the usual standard model backgrounds, events
are removed in which the invariant mass of the opposite-sign same-flavor
(OSSF) leptons is more than 75 GeV.  CMS then performs a neural net analysis
with 7 variables to suppress the backgrounds from SM production of $Z+$jets
and virtual photon$+$jets (called Drell Yan in their terminology) to achieve a
signal to background ratio $S/B$ of $1/3$.  There are not enough details or
referenced notes regarding the neural net for us to reproduce the results, but
we can confirm agreement for the backgrounds considered.  However, the new
backgrounds calculated here far outweigh the ones studied by CMS.  As all of
our estimated backgrounds are larger than those determined by CMS, we explain
in detail where our simulations agree and where and why they differ.

There are three overriding observations to keep in mind.  First, each
significant background involving leptons from heavy flavor decays, not
estimated by CMS, is more than 10 times larger than the largest background
considered by CMS.  Second, as explained below, our first benchmark process,
$WZ$ includes additional physics that enhances this background.  Third, the
ISASUGRA 7.69 evolution code used by CMS for the supersymmetric spectrum gives
mass differences and branching fractions incompatible with the more recent
version employed in our work.  Nevertheless, we find good agreement with the
CMS results when we consider only their backgrounds and use their assumptions.
Therefore, we are confident that the simulation improvements discussed below
are valid and will be useful for future studies.

\subsection{$WZ$, $t \bar{t}$, and $Wt$ final states}

In order to maintain consistency across processes considered, we elect to
present results using leading-order cross sections rather than introducing
next-to-leading order (NLO) $K$ factors.  While NLO calculations exist for
most of these processes~\cite{NLO}, there is considerable overlap at NLO
between final states, e.g., $bZ$ and $b\bar bZ$.  Furthermore, the jet veto
will reject some of the higher order radiation that enhances the cross
sections.  Proper matching between orders is beyond the scope of this paper.

Since our goal is to establish a direct connection with the CMS analysis, we
first confirm that our detector simulation reproduces the CMS detector
simulation for reconstructed leptons.  Performing a PYTHIA-based calculation
of $WZ$, minus decays to taus, and using the CMS $K$-factor of 2, we predict
171 events after cuts in 30 \fbi\ vs.\ 173 events in Table~6 of the CMS
analysis~\cite{CMSpaper}.  Any differences in lepton reconstruction enter 3
times; hence, this comparison shows that our detector simulation agrees
exceedingly well with the CMS detector simulation.

\begin{figure}[htb]
\centering
\includegraphics[width=3.25in]{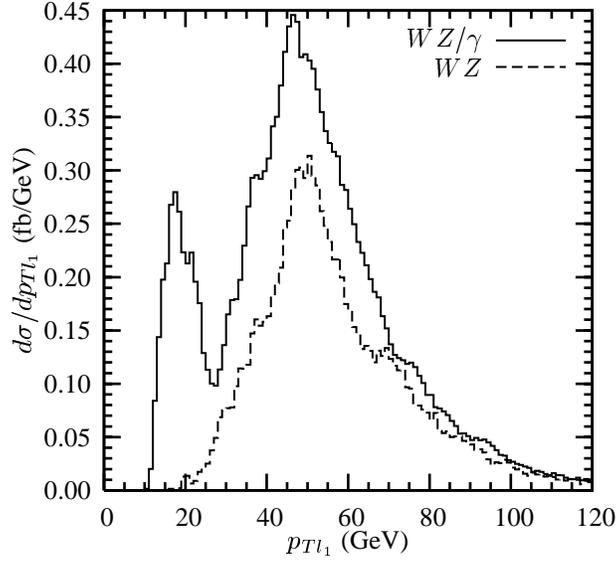}
\caption{Transverse momentum ($p_T$) distribution of the leading-$p_T$ lepton
in $WZ/\gamma$ compared with $WZ$.
\label{fig:ptlo}}
\end{figure}

\begin{figure}[htb]
\centering
\includegraphics[width=3.25in]{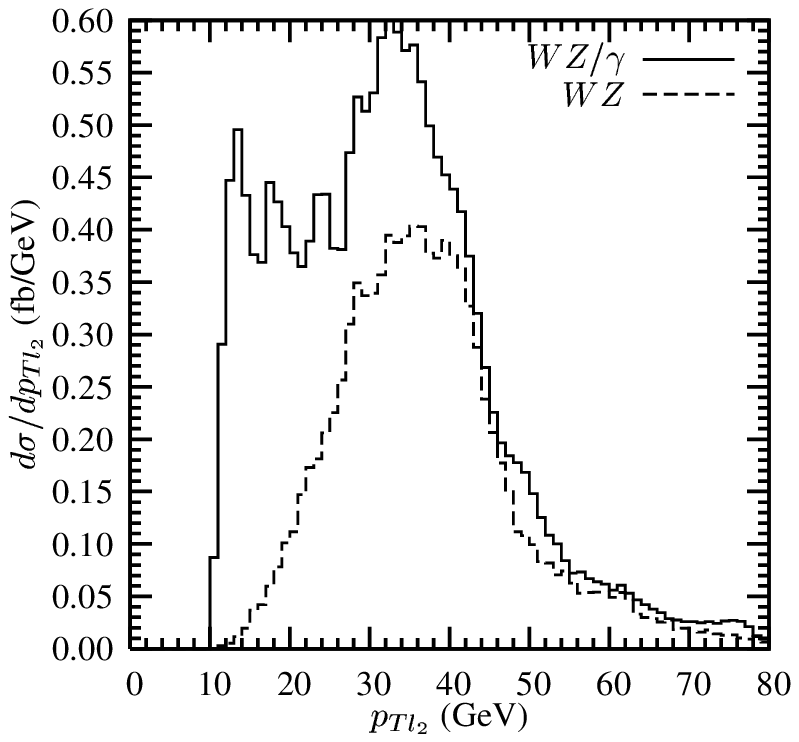}
\caption{Transverse momentum ($p_T$) distribution of the second leading-$p_T$
lepton in $WZ/\gamma$ compared with $WZ$.
\label{fig:ptltw}}
\end{figure}

\begin{figure}[htb]
\centering
\includegraphics[width=3.25in]{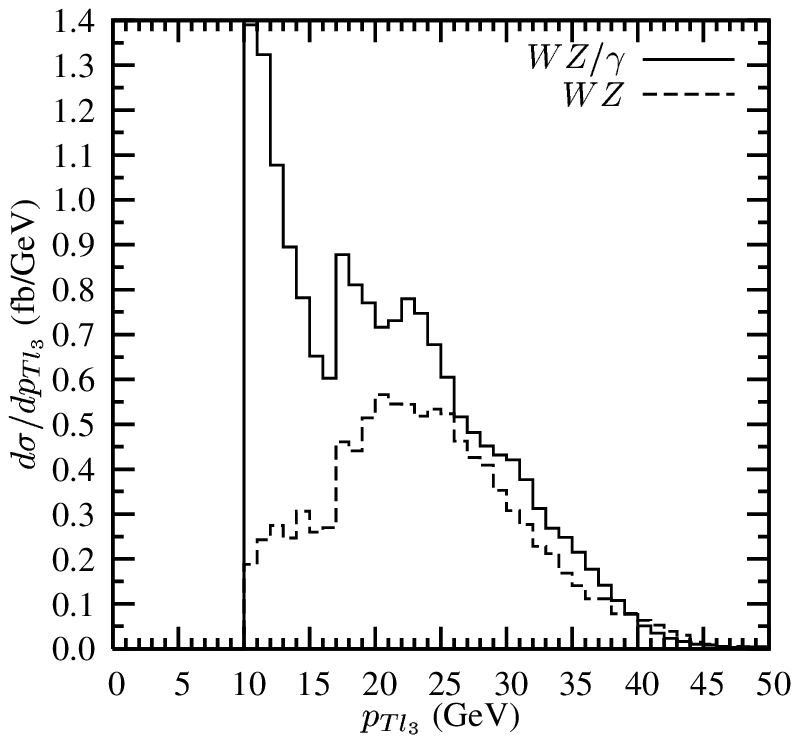}
\caption{Transverse momentum ($p_T$) distribution of the lowest-$p_T$ lepton
in $WZ/\gamma$ compared with $WZ$.
\label{fig:ptlth}}
\end{figure}

The PYTHIA code used by CMS for $WZ$ production omits virtual photon
$\gamma^{*}$ production and interference with the $Z$.  Given that the $Z$
peak will be removed by cuts, we find that the additional contribution from
$W$ plus a virtual photon is at least as important.  This result agrees with
other work~\cite{backgrounds,Campbell:1999ah} in which the importance of the
virtual photons is emphasized.  In all of our simulations we include the
virtual photon continuum and photon-$Z$ interference along with the $Z$.  In
the case of the SM $WZ$ contribution, we denote the full contribution from
$WZ$, $W \gamma^{*}$, and $W+Z , \gamma^{*}$ interference by $WZ/\gamma$.  In
Figs.\ \ref{fig:ptlo}--\ref{fig:ptlth} we show the transverse momentum
distribution of the $p_T$-ordered leptons from $WZ/\gamma$ compared with the
distributions from $WZ$ alone.  The peak at 20 GeV in Fig.\ \ref{fig:ptlo} is
produced by the minimum $p_T$ cuts on the individual leptons that make up the
virtual photon contribution.  Essentially all of the leptons from off-shell
photons combine survive the $Z$-peak mass cut.  When there are 3 muons or 3
electrons, both OSSF pairings pass the cuts.  Hence, about $1/2$ of the
virtual photon events appear twice when counting events.

Table \ref{tab:CMScomp} compares the LO $WZ/\gamma$ cross section with a
virtual photon included to a CMS-like NLO $WZ$ estimate using PYTHIA.  Given
that a $K$-factor of 2 is used in the CMS estimate, there is a full factor of
6 difference between $WZ$ built into PYTHIA, and $WZ/\gamma$ generated by
MadEvent and showered by PYTHIA.  Nearly $1/2$ of the excess is attributable
to the virtual photon that is not included in PYTHIA.  Some of the excess
comes from correlations captured by the full matrix element that are sensitive
to the cuts.  In particular, angular correlations among the final-state
leptons are missing from the CMS study.  In order to study these correlations,
and to quote a more realistic background, we use the full matrix element for
$WZ/\gamma\to lll\nu$ from MadEvent showered through PYTHIA in our results
below.

\begin{table}[ht]
\caption{Comparison of two backgrounds calculated by CMS and by
this study.  $WZ/\gamma$ and $t\bar t$ are our estimates.  $WZ_\mathrm{PYT}$
is calculated from PYTHIA and normalized to the CMS NLO total cross section.
$t\bar t_{\mathrm{CMS}\,j}$ uses the CMS acceptance for the jet veto, based on
a matched calculation that produces significantly more radiation than a
leading order calculation.  The second column shows the number of events
expected with 30 fb$^{-1}$ of integrated luminosity after the requirements
that there be 3 leptons and no jets with $E_{Tj} > 30$~GeV.  In the third
column an additional requirement is imposed that the invariant mass of any
pair of OSSF leptons be no greater than $75$~GeV.
\label{tab:CMScomp}}
\begin{tabular}{lD{.}{.}{2}D{.}{.}{2}}
& \multicolumn{1}{c}{$N^l=3$,} & \\
Channel &\multicolumn{1}{c}{no jets}&
\multicolumn{1}{c}{$M_{ll}^\mathrm{OSSF}<75$~GeV} \\ \hline
$WZ/\gamma^\mathrm{LO}$ & 1880 & 538 \\
$t\bar t$ & 1540 & 814 \\ \hline
$WZ_\mathrm{PYT}^\mathrm{NLO}$ & 661 & 171 \\
$t\bar t_{\mathrm{CMS}\,j}$ & 394 & 208
\end{tabular}
\end{table}

The remaining increase in the normalization of the cross section is connected
with the requirement that there be no jets with $E_T > 30$~GeV.  We compare
$WZ\to lll\nu$ events generated with MadEvent and showered with PYTHIA to $WZ$
events which are generated, decayed, and showered all within PYTHIA.  In the
process we observe that the initial-state QCD radiation spectrum is harder in
the all-PYTHIA events.  The jet veto rejects a larger fraction of this pure
electroweak process assuming PYTHIA generation, and hence leads to a smaller
normalization of the cross section after cuts.  This effect is consistent
across all processes we calculate, and hence leads to an overall uncertainty
in the absolute cross section.  Our main concern in this paper is to
understand the \textit{relative} importance of the various backgrounds, and
their shapes.  When data are in hand, we expect the total background to be
normalized to data.  Because most of the processes we evaluate are not in
PYTHIA, we perform and demonstrate the cross sections for MadEvent generated
events, and discuss the effect of normalization on a discovery signal at the
end.

CMS also quotes their estimated background from $t\bar t$ production.  It is
important to compare with their result because this background requires one
lepton from a $b$ decay.  The CMS analysis uses a matched
\cite{Slabospitsky:2002ag} $t\bar t$ cross section which produces more jets in
the final state than our LO calculation.  Hence, CMS has a higher rejection
rate based on their jet veto.  If we rescale our result using the efficiency
of their jet veto instead of ours, we find 208 events in 30 \fbi\ vs.\ 239
events in the CMS paper.  We find excellent agreement between our estimate of
a background with a lepton from a $b$ and the CMS estimate.  A similar
rescaling for $Wt$ would give us 37 events vs.\ 45 in the CMS paper.  Our
systematic 10\% underestimate of the effect of isolated leptons from $b$ decay
in the $t \bar{t}$ sample suggests our other results involving leptons from
heavy flavor decays are conservative.  The comparison between our calculation
of $t\bar t$ and the jet-acceptance reduced version are summarized in Table
\ref{tab:CMScomp}.

\subsection{Heavy flavor final states}
\label{sec:hffs}

Before addressing the importance of leptons from heavy-flavor decays, we
comment on one subtlety in the supersymmetric models studied.  The 243 events
we find for SUSY point LM9 are similar to the 238 events expected by CMS.
However, we predict 123 events vs.\ 91 for point LM7, and 44 events vs.\ 70
events for LM1.  The differences are attributable to the use of ISASUGRA 7.75
in our study vs.\ ISASUGRA 7.69 by CMS\footnote{We are unable to use ISASUGRA
7.69 due to a change in data format.}.  There was a small bug in the evolution
codes in ISASUGRA 7.69 that, when corrected, leads to 10 GeV shifts in the
neutralino mass spectrum.  The $Z$ mass peak cut is sensitive to the endpoint
of the lepton spectrum, $M_{ll}^\mathrm{max} = \sqrt{(m^2_{\Zb} -
m^2_{\Slp})(m^2_{\Slp} - m^2_{\Za})/m^2_{\Slp}}$ for the three-body decays at
LM1, and $M_{ll}^\mathrm{max} = m_{\Zb}-m_{\Za}$ for LM7 and LM9.  The
endpoint for LM1 is just above the $Z$ mass-peak cut, and the results are
sensitive to small shifts in the neutralino masses.  Additionally, branching
fractions for LM7 change between ISASUGRA versions as $\tau$'s are predicted
to become more or less important.  Therefore, only point LM9 is directly
comparable to the CMS study we discuss here, but we include the other points
as future studies will likely use the newer evolution formulae.

In the previous subsection, we establish agreement in three cases in which we
can compare directly with CMS.  In this subsection, we turn to our main
results.  In Table \ref{tab:CMSres} we show the number of signal and
background events expected with 30 \fbi\ in our CMS-like analysis.  For each
subprocess we list the number of trilepton events expected after the jet veto
is applied as well as the number of events that survive the $Z$ peak mass cut.
We also tabulate the number of events that survive after two additional cuts
which we describe below.

\begin{table}[ht]
\caption{Leading order signal and background statistics for several final
states in a CMS-like analysis, and with two additional cuts.  The second
column shows the number of events expected with 30 fb$^{-1}$ of integrated
luminosity after the requirements that there be 3 isolated leptons and no jets
with $E_{Tj} > 30$~GeV.  In the third column an additional requirement is
imposed that the invariant mass of any pair of OSSF leptons be no greater than
$75$~GeV.  The virtual photon components are large after cuts.
\label{tab:CMSres}}
\begin{tabular}{lD{.}{.}{2}D{.}{.}{2}D{.}{.}{2}D{.}{.}{2}}
& \multicolumn{1}{c}{$N^l=3$}& & & \multicolumn{1}{c}{Angular} \\
Channel &\multicolumn{1}{c}{no jets}&
\multicolumn{1}{c}{$M_{ll}^\mathrm{OSSF}<75$~GeV} &
\multicolumn{1}{c}{${\MET > 30}\textrm{ GeV}$} &
\multicolumn{1}{c}{cuts} \\ \hline
LM9 & 248 & 243 & 160 & 150 \\
LM7 & 126 & 123 & 89 & 85 \\
LM1 & 46 & 44 & 33 & 32 \\ \hline
$WZ/\gamma$ & 1880 & 538 & 325 & 302 \\
$t\bar t$ & 1540 & 814 & 696 & 672 \\
$tW$ & 273 & 146 & 123 & 121 \\
$t\bar b$ & 1.1 & 1.0 & 0.77 & 0.73 \\
$bZ/\gamma$ & 14000 & 6870 & 270 & 177 \\
$cZ/\gamma$ & 3450 & 1400 & 45 & 35 \\
$b\bar bZ/\gamma$ & 8990 & 2220 & 119 & 103 \\
$c\bar cZ/\gamma$ & 4680 & 1830 & 69 & 35 \\
$b\bar bW$ & 9.1 & 7.6 & 5.6 & 5.3 \\
$c\bar cW$ & 0.19 & 0.15  & 0.12 & 0.11
\end{tabular}
\end{table}

\begin{figure}[htb]
\centering
\includegraphics[width=3.25in]{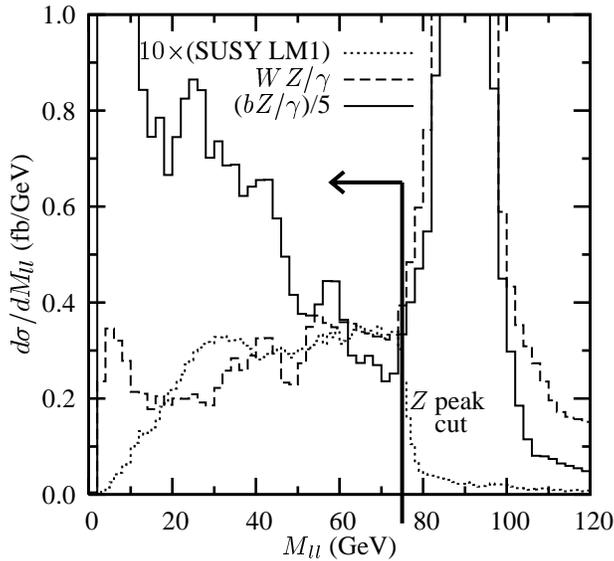}
\caption{Invariant mass of the opposite-sign same-flavor (OSSF) lepton pairs
for the SUSY LM1 signal ($\times 10$, dotted), the $WZ/\gamma$ background
(dashed), and the $bZ/\gamma$ background (divided by 5, solid).  Other
$Z/\gamma+$heavy flavor backgrounds (not shown) have the same shape as
$bZ/\gamma$.
\label{fig:MllLMo}}
\end{figure}

\begin{figure}[htb]
\centering
\includegraphics[width=3.25in]{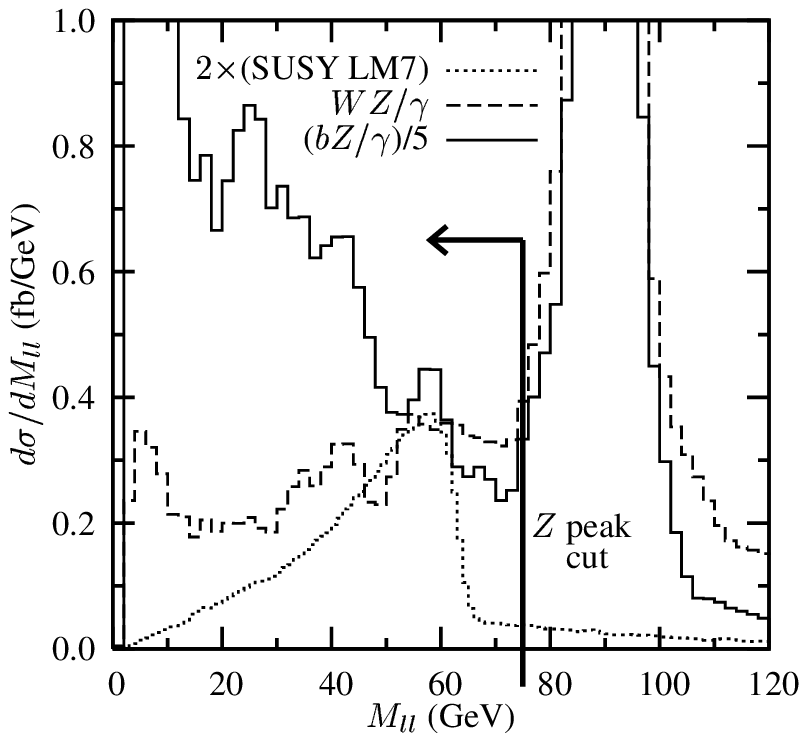}
\caption{Invariant mass of the opposite-sign same-flavor (OSSF) lepton pairs
for the SUSY LM7 signal ($\times 2$, dotted), the $WZ/\gamma$ background
(dashed), and the $bZ/\gamma$ background (divided by 5, solid).  Other
$Z/\gamma+$heavy flavor backgrounds (not shown) have the same shape as
$bZ/\gamma$.
\label{fig:MllLMs}}
\end{figure}

\begin{figure}[htb]
\centering
\includegraphics[width=3.25in]{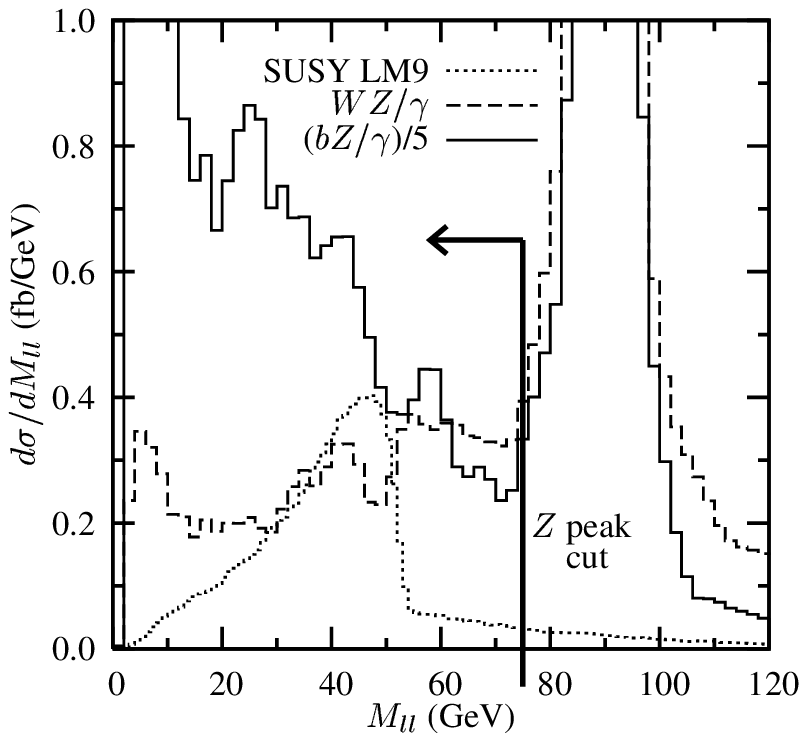}
\caption{Invariant mass of the opposite-sign same-flavor (OSSF) lepton pairs
for the SUSY LM9 signal (dotted), the $WZ/\gamma$ background (dashed), and the
$bZ/\gamma$ background (divided by 5, solid).  Other $Z/\gamma+$heavy flavor
backgrounds (not shown) have the same shape as $bZ/\gamma$.
\label{fig:MllLMn}}
\end{figure}

Addressing the contributions of $Z/\gamma$ plus heavy flavors and $W$ plus
heavy flavors, we see in Table \ref{tab:CMSres} that, before the $Z$ peak cut,
$Z/\gamma$ plus heavy flavors produces trileptons 16 times more often than $W
Z/\gamma$.  After the $Z$ peak cut, the ratio rises to 23 times $WZ/\gamma$.
In particular, $b\bar bZ/\gamma$, which includes a virtual photon, is over 30
times the CMS estimate of the cross section that does not have a photon.  In
Figs.\ \ref{fig:MllLMo}--\ref{fig:MllLMn}, we see the opposite-sign
same-flavor (OSSF) dilepton invariant mass for the signals, $WZ/\gamma$
background, and $(bZ/\gamma)/5$.  Cutting out the $Z$ mass peak reduces the
backgrounds, but the remaining tail for $Z/\gamma$ plus heavy flavors
overwhelms the signal.

The number of handles available to reject the huge background from
$Z/\gamma+$heavy flavors is limited.  In Ref.\ \cite{Sullivan:2006hb} we
recommend raising the minimum lepton $p_T$ threshold since the lepton $p_T$
spectrum from $b$ and $c$ decays tends to fall rapidly.  In typical trilepton
studies, however, the leptons are very soft.  Hence, any increase in the cut
on the lepton $p_T$ tends to reject too much of the signal.

\begin{figure}[htb]
\centering
\includegraphics[width=3.25in]{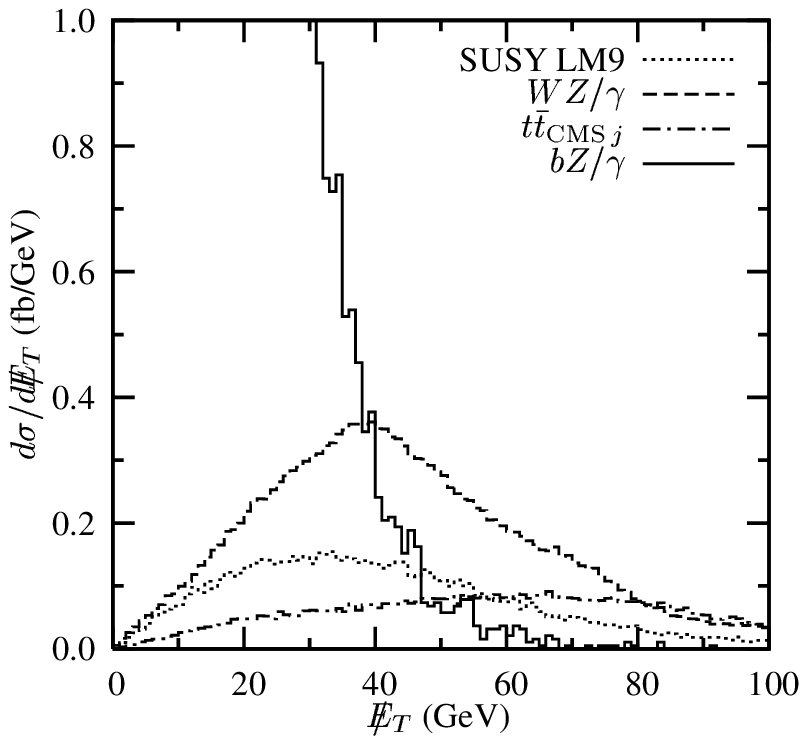}
\caption{Missing transverse energy spectrum $\MET$ of the opposite-sign
same-flavor (OSSF) lepton pairs for the SUSY LM9 signal (dotted), the
$WZ/\gamma$ background (dashed), $t\bar t$ (dot-dashed), and $bZ/\gamma$
(solid).  Other $Z+$heavy flavor backgrounds (not shown) have the same shape
as $bZ/\gamma$.
\label{fig:METLMn}}
\end{figure}

Missing transverse energy $\MET$ is a partial discriminator.  The SUSY signals
contain invisible neutralinos which leave a broad range of $\MET$ in the
detector.  In Fig.\ \ref{fig:METLMn} we show the $\MET$ spectrum for the SUSY
LM9 signal and for the $WZ/\gamma$ and $bZ/\gamma$ backgrounds.  Trilepton
signatures from $t\bar t$ production generally have two neutrinos which lead
to large missing energy.  The contribution from $Z/\gamma+$heavy flavor
processes peaks at over 400 times the size of the LM9 signal at low $\MET$,
but it falls rapidly to below the signal by $\MET>50$ GeV.  These differences
present both opportunities and challenges, especially since the precision of
$\MET$ measurements is not as great as one might prefer.

In the fourth column of Table \ref{tab:CMSres}, we show that the requirement
$\MET>30$ GeV removes most of the $Z/\gamma+$heavy flavor
backgrounds\footnote{We do not estimate the $Z/\gamma+$light jets background
here because it requires an accurate prediction of the rate for jets to fake
leptons.  However, $Z/\gamma+$light jets has a nearly identical $\MET$
spectrum.  Hence, a $\MET$ cut should perform similarly well.} for a modest
loss of signal.  A cut below 20 GeV is not as useful and is likely not
achievable at the LHC.  A cut above 40 GeV removes most of the $Z/\gamma+X$
backgrounds, but it begins to significantly reduce the signal and is of little
additional help with $WZ/\gamma$ and $t\bar t$ backgrounds.  The sharply
falling $\MET$ spectrum in $Z/\gamma+X$ is extremely sensitive to
uncertainties in the measurement of $\MET$.  This uncertainty makes it
difficult to predict absolute cross sections after cuts.  On the other hand,
this sensitivity could provide an excellent opportunity to \textit{measure the
background in situ} and reduce concerns regarding modeling details.  The
background can be fit in the data and the $\MET$ cut adjusted to optimize the
purity of the sample.

\begin{figure}[htb]
\centering
\includegraphics[width=3.25in]{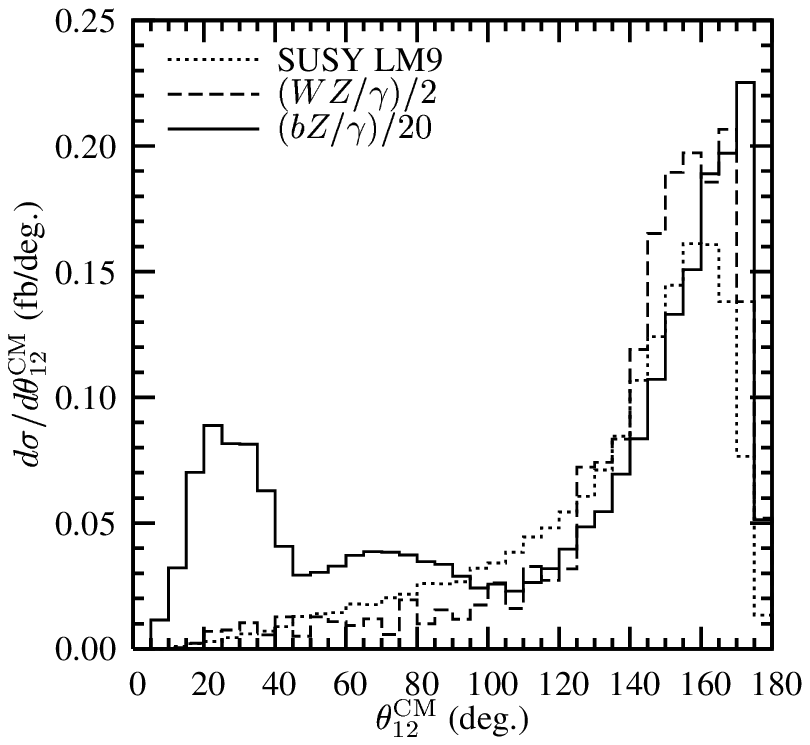}
\caption{Angular distribution between the leading $p_T$-ordered leptons in the
trilepton center-of-momentum frame.  $t\bar t$ is nearly identical in shape to
SUSY LM9, and the $Z/\gamma+X$ backgrounds are similar to $bZ/\gamma$.
\label{fig:ThetaA}}
\end{figure}

\begin{figure}[htb]
\centering
\includegraphics[width=3.25in]{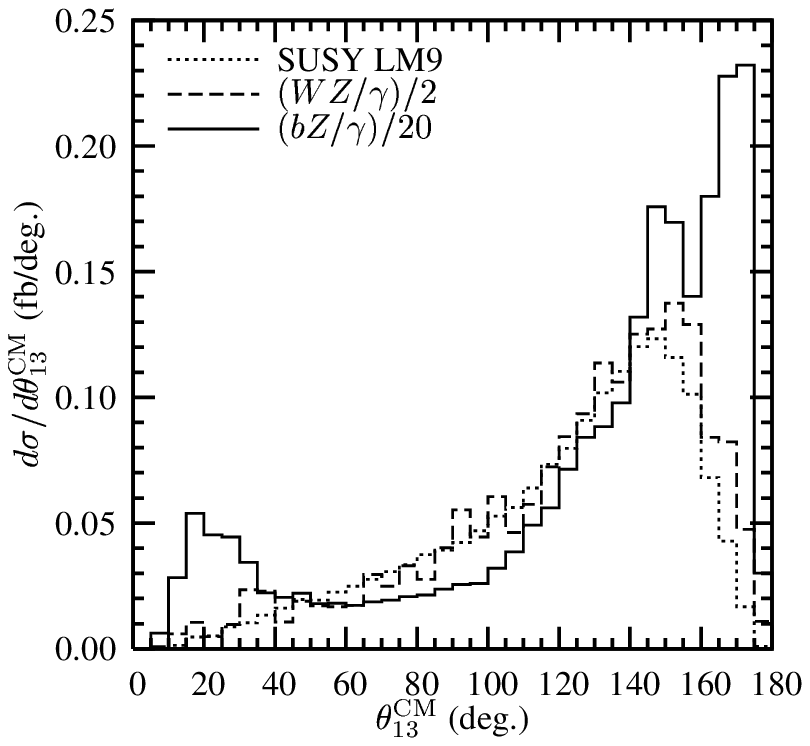}
\caption{Angular distribution between the leading and third $p_T$-ordered
lepton in the trilepton center-of-momentum frame.  $t\bar t$ is nearly
identical in shape to SUSY LM9, and the $Z/\gamma+X$ backgrounds are similar
to $bZ/\gamma$.
\label{fig:ThetaB}}
\end{figure}

\begin{figure}[htb]
\centering
\includegraphics[width=3.25in]{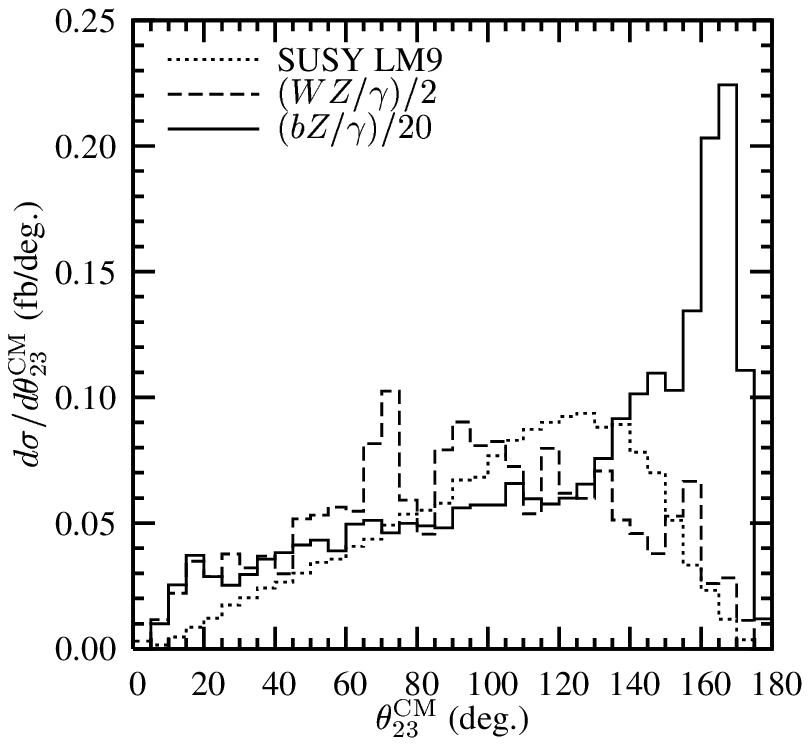}
\caption{Angular distribution between the second and third $p_T$-ordered
lepton in the trilepton center-of-momentum frame.  $t\bar t$ is nearly
identical in shape to SUSY LM9, and the $Z/\gamma+X$ backgrounds are similar
to $bZ/\gamma$.
\label{fig:ThetaC}}
\end{figure}

Since the accuracy of $\MET$ measurements is limited, we examine instead the
utility of angular cuts without a $\MET$ cut.  There are significant angular
correlations in the $Z/\gamma+$heavy flavor backgrounds that are different
from those in the SUSY trilepton signals or the $WZ/\gamma$ and $t\bar t$
backgrounds.  In Figs.\ \ref{fig:ThetaA}--\ref{fig:ThetaC} we plot the angular
distribution $\theta_{ij}^{\mathrm{CM}}$ between pairs of $p_T$-ordered
leptons in the trilepton center-of-momentum (CM) frame without a $\MET$ cut.
The $Z/\gamma+$heavy flavor backgrounds have significant peaks at both small
and large angles.  The signal and other backgrounds either peak only at large
angles ($\theta^\mathrm{CM}_{12}$, $\theta^\mathrm{CM}_{13}$), or are fairly
central ($\theta^\mathrm{CM}_{23}$).

We examine the impact of these angular correlations by themselves by imposing
three angular cuts in the trilepton center-of-momentum frame:
$\theta^\mathrm{CM}_{12}>45^\circ$, $\theta^\mathrm{CM}_{13}>40^\circ$, and
$\theta^\mathrm{CM}_{23}<160^\circ$.  These angle cuts reduce the heavy flavor
backgrounds and $WZ/\gamma$ by $\sim 30\%$, with only a 5\% reduction of the
signal.  These cuts could be further optimized, but in general they are more
useful for increasing purity than for increasing significance.

\begin{figure}[htb]
\centering
\includegraphics[width=3.25in]{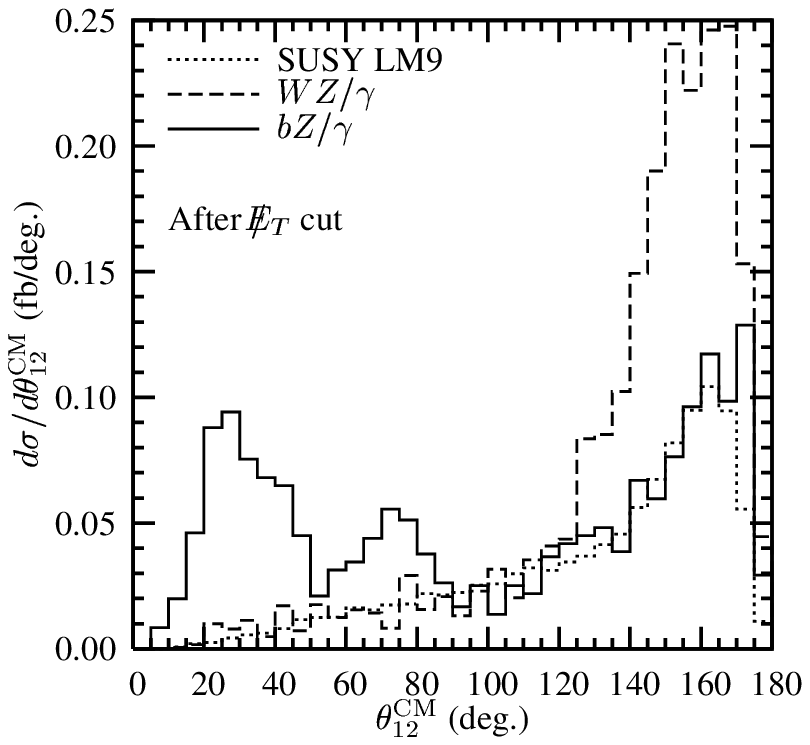}
\caption{Angular distribution between the leading $p_T$-ordered leptons in the
trilepton center-of-momentum frame, after a missing energy cut $\MET>30$ GeV.
$t\bar t$ is nearly identical in shape to SUSY LM9, and the $Z/\gamma+X$
backgrounds are similar to $bZ/\gamma$.
\label{fig:FThetaA}}
\end{figure}

\begin{figure}[htb]
\centering
\includegraphics[width=3.25in]{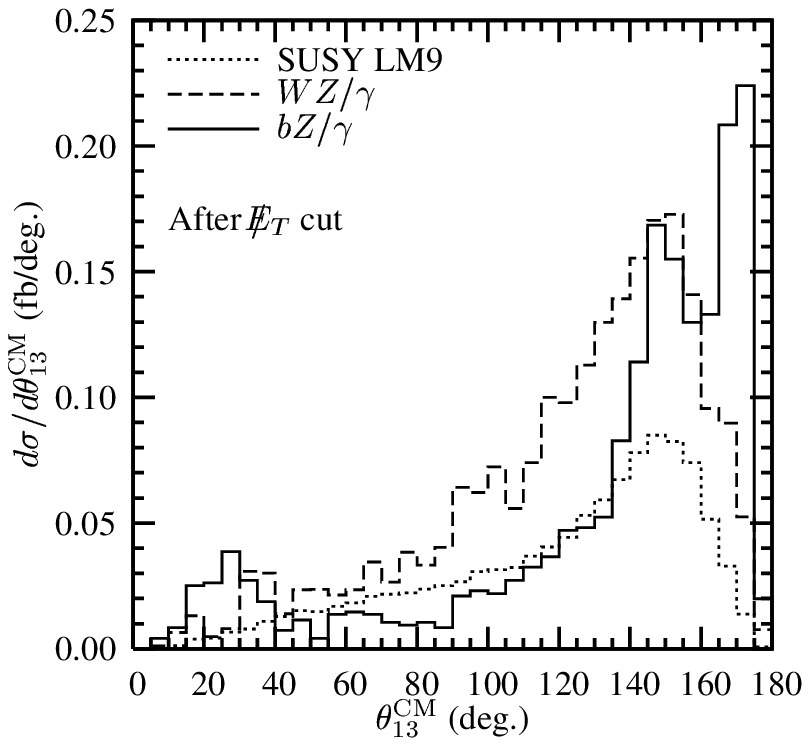}
\caption{Angular distribution between the leading and third $p_T$-ordered
lepton in the trilepton center-of-momentum frame, after a missing energy cut
$\MET>30$ GeV.  $t\bar t$ is nearly identical in shape to SUSY LM9, and the
$Z/\gamma+X$ backgrounds are similar to $bZ/\gamma$.
\label{fig:FThetaB}}
\end{figure}

\begin{figure}[htb]
\centering
\includegraphics[width=3.25in]{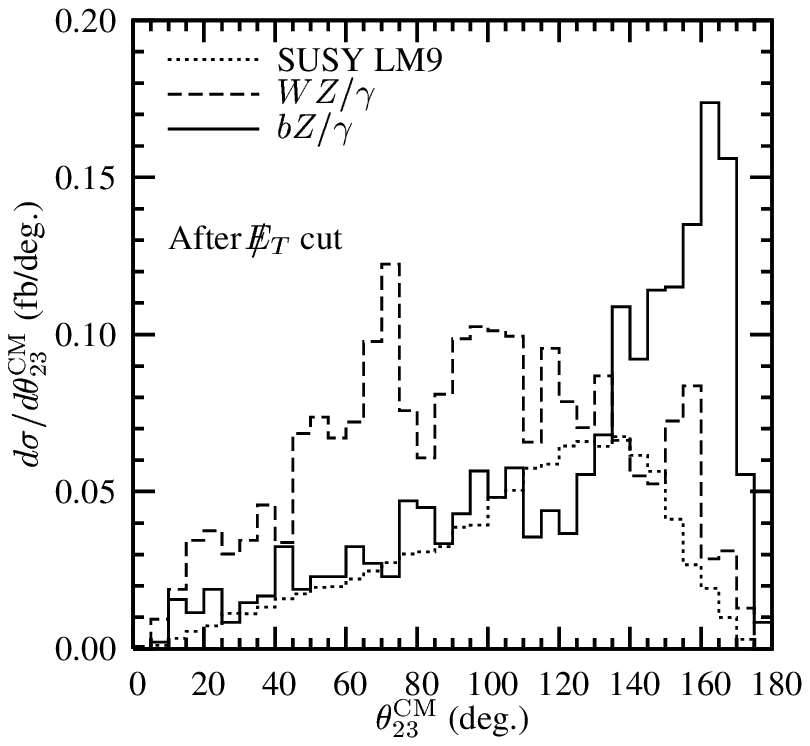}
\caption{Angular distribution between the second and third $p_T$-ordered
lepton in the trilepton center-of-momentum frame, after a missing energy cut
$\MET>30$ GeV.  $t\bar t$ is nearly identical in shape to SUSY LM9, and the
$Z/\gamma+X$ backgrounds are similar to $bZ/\gamma$.
\label{fig:FThetaC}}
\end{figure}

Figures \ref{fig:FThetaA}--\ref{fig:FThetaC} show the angular distributions
between pairs of $p_T$-ordered leptons in the trilepton CM frame after the
missing transverse energy cut.  The last column of Table~\ref{tab:CMSres}
demonstrates the effect of adding the three angular cuts after the $\MET$ cut.
There is almost no correlation between the effects of the angle cuts and the
missing transverse energy cut except in $bZ/\gamma$, where the cut is slightly
more effective after the $\MET$ cut (acceptance is 0.32 vs.\ 0.37).  The real
advantage of angular cuts over a $\MET$ cut is that the angles are well
measured.  Whether or not missing transverse energy can be well-measured, the
use of the angular cuts will improve any final analysis.

One class of backgrounds not treated so far is one in which all three leptons
come from heavy flavor production ($b\bar bb\bar b$, $c\bar cc\bar c$, $b\bar
bc\bar c$).  As mentioned in Sec.\ \ref{sec:sim}, practical limitations
prevent us from simulating this background directly.  Instead we provide
rough estimates for the trilepton signature from three sources of $b\bar
bb\bar b$ production.  First, using $Wb\bar b$ production, we obtain an
estimate for the probability of finding an isolated lepton at the LHC of
$7.5\times 10^{-3}$ per $b$.  The cross section for direct production of
$b\bar bb\bar b$ from MadEvent multiplied by $(7.5\times 10^{-3})^3$ gives
about 500 trilepton events in 30 \fbi\ of data.  Second, multiple interactions
per beam crossing, where more than one $b\bar b$ pair is produced, can lead to
a trilepton signature.  For a total inelastic cross section of about 80
mb~\cite{CMSTDR2}, the rate for two scatters to give three isolated leptons is
about 60 events per interaction per 30 \fbi.\footnote{Multiply the ratio of
$b\bar b$ production to total inelastic cross section times the $b\bar b$
cross section.}  At 10 interactions per crossing, this method leads to about
600 events, though the jet veto will likely reduce this figure somewhat.

Multiple scattering in a given interaction (e.g., double parton scattering),
or even showering of heavy quarks, is a third source, arising from successive
production of $b\bar b$ pairs.  The rate may be comparable to direct 4-$b$
production after cuts on transverse momentum~\cite{DelFabbro:2002pw}, and it
can double the backgrounds listed so far.  These rough estimates are the same
order of magnitude as the number of events from $Z/\gamma+$heavy flavor
processes, and they are a potentially serious problem with the default
analysis.  A sum over all production processes and decays to both muons and
electrons would increase these numbers by an order of magnitude.  On the other
hand, in Ref.\ \cite{Sullivan:2006hb} we observe that a missing energy cut
serves to significantly reduce the net cross section from pure QCD processes.
Even if the suppression is only as strong as for $Z/\gamma+$heavy flavors, the
background can and should be measured \textit{in situ} as a function of
$\MET$, and then removed by adjusting the $\MET$ cut as needed.

Unlike the CMS study in Ref.\ \cite{CMSpaper}, we find that the $E_T > 30$~GeV
jet veto and the $Z$ mass-peak cut are not sufficient to find more than a
$2\sigma$ evidence for SUSY point LM9 (the most optimistic).  However, the
addition of a cut on $\MET$ is sufficient to find at least a $4\sigma$ excess
(and perhaps an $8\sigma$ excess if a CMS-like initial-state radiation
estimate is used).  In Table \ref{tab:CMSsig}, we summarize our expected
sensitivity in 30 \fbi\ for point LM9.  We also estimate the sensitivity after
the background from fakes is included (using fake rates from Table~6 of the
CMS study).  The background from fakes has little effect on the final
analyses.  The angular cuts mentioned above do not appear to influence the
significance of the signal, but we emphasize again that angles will be much
better measured than missing transverse energy and may be useful in
compensating for experimental uncertainties.  The other SUSY points studied do
less well with any cuts.  Using the more optimistic CMS-like jet veto scenario
and Tables \ref{tab:CMSres} and \ref{tab:CMSsig}, we find the best
significance for points LM7 and LM1 to be $4.2\sigma$ and $1.6\sigma$,
respectively, in 30 \fbi\ of integrated luminosity.

\begin{table}[ht]
\caption{The significance for SUSY point LM9 is listed for each cut, as well
as the significance assuming all background events are reduced to the CMS jet
veto acceptance (a reduction factor of 3.6).  In parentheses are estimates of
the total significance assuming additional detector backgrounds from Table~6
of the CMS analysis not directly evaluated in this study.
\label{tab:CMSsig}}
\begin{tabular}{lcccc}
 & ${N^l=3}$ & & & Angular \\
 &\multicolumn{1}{c}{no jets}&
$M_{ll}^\mathrm{OSSF}<75$~GeV &
${\MET > 30}\textrm{ GeV}$ &
cuts \\ \hline
$S/\sqrt{B}_\mathrm{LM9}$ & 1.33 & 2.07(1.79) & 3.93(3.74) & 3.94(3.79) \\
$S/\sqrt{B}_\mathrm{LM9}^{\mathrm{CMS}\,j}$ & 2.63 & 4.09(3.54) & 7.78(7.39)
& 7.79(7.49)
\end{tabular}
\end{table}

We conclude this section with a note regarding flavor and sign combinations.
For the SUSY trilepton signals and standard model backgrounds we consider,
there in no difference in production rate of the states $\mu\mu\mu$,
$e\mu\mu$, $ee\mu$, or $eee$.  Any observed difference is expected to come
from the slightly larger acceptance for isolated electrons over muons from
heavy-quark decays, balanced against the slightly larger rejection of electron
events over muons from the $Z$-peak mass cut.  This fact may be useful as a
check of detector acceptances, but it has no power to resolve the signal.  The
$W^\pm Z/\gamma$, $tW^\pm$, $b\bar bW^\pm$, and $c\bar cW^\pm$ cross sections
will produce a slight enhancement of 2 positive leptons over 2 negative
leptons since parton distribution functions for protons favor $W^+$ production
over $W^-$ production (e.g., $W^+Z/\gamma$:$W^-Z/\gamma \approx 3$:$2$).  It
may be possible to use this sign difference to constrain the total $W+X$
background in this sample \textit{in situ}, and improve the control of some
systematic errors.  Overall, this type of analysis may be useful for discovery
of BSM trilepton signatures that favor the production of muons over electrons
(or vice versa).

\section{Comparison to ATLAS}
\label{sec:ATLAS}

The ATLAS Collaboration has considered the search for supersymmetry in the
trilepton channel for a long time \cite{ATL97}.  More recently, the ATLAS
Collaboration has been utilizing trilepton signatures from supersymmetry to
study variations on electron identification and jet rejection algorithms.  We
examine trileptons in the context of the SU2 test point as examined in
preliminary contributions to the ATLAS CSC 7 Note \cite{ATLANAL}.  The SU2
test point is in the ``focus point'' region of mSUGRA parameter space, and its
parameters are $m_0=3550$ GeV, $m_{1/2}=300$ GeV, $A_0=0$ GeV, $\tan\beta=10$,
and $\mu>0$.

One subtlety is that the SUSY evolution in ISASUGRA 7.75 predicts that if the
top-quark mass is less than 175 GeV, these mSUGRA parameters fail to break
electroweak symmetry.  The lightest neutralino mass is extremely sensitive to
the top-quark mass here, e.g., $m_{\Za}$ changes by 20 GeV with a 100 MeV
shift in the top-quark mass.  Hence, we tune the top-quark mass to 175.1 GeV
in order to reproduce the mass spectrum used by ATLAS as closely as possible.
Ultimately the specific model is not important, but rather we desire an
estimate of how small a cross section can be observed in the general analysis
given the standard model backgrounds.

The basic ATLAS analysis \cite{ATLANAL} is nearly identical to that used by
CMS, with a few modifications we indicate here.  They first demand 3 isolated
leptons (see the Appendix), where $p_{T\mu}>10$ GeV and $p_{Te}>15$ GeV, and
they veto events with jets having $E_{Tj} > 20$ GeV.  The default analysis
then invokes a maximum separation cut between OSSF leptons of $\Delta
R_{ll}<2.6$, to emphasize the region of phase space in which the signal is
most significant.  The $Z$ peak is removed by cutting out the OSSF invariant
mass region 80 GeV $<M_{ll}^\mathrm{OSSF} <$ 100 GeV; and a modest amount of
missing transverse energy is required $\MET > 10$ GeV.

We begin by comparing how well we reproduce the ATLAS expectations for the
signal.  For the SU2 fixed point, ATLAS considers the simultaneous production
of $\widetilde{\chi}_i^0\widetilde{\chi}_j^\pm$, where $i=2$--4, and $j=1, 2$.
For this analysis we examine only $\Zb \Wpa$, as was done for CMS.  For
30~\fbi\ of integrated luminosity, the ATLAS study expects 37 events after
cuts from $\widetilde{\chi}_2^0\widetilde{\chi}_1^\pm$ alone.  We obtain
excellent agreement with 37 events after cuts once we scale up the production
cross section by $1.28$ to match the input cross section used by ATLAS.
Rather than list results scaled to older analyses, we present our results
below using the newer ISASUGRA 7.75 spectrum.

Our acceptance for $WZ$ agrees well at each cut level with the PYTHIA-based
ATLAS analysis we are following.  However, the missing virtual photon
contribution to trileptons from $W\gamma^*$ is again important to the total
number of backgrounds events.  Our estimate of 466 events from $t\bar t$
agrees well the 453 events expected by the ATLAS study.  This adds
confirmation that we reproduce ATLAS expectations for leptons, jets, and
$\MET$ even after cuts.

In Table \ref{tab:ATLres} we summarize our results for the ATLAS-like study of
the SU2 focus point.  At the level of three leptons and no jets, and after the
default ATLAS cuts, the trilepton signature from heavy flavor decays plus a
$Z$ or virtual photon exceeds the other standard model backgrounds by at least
a factor of 10.  The weak missing transverse energy cut ${\MET}>10$ GeV is not
sufficient to effectively reduce the background from $Z/\gamma+$heavy flavor
decays.  Raising the cut on missing transverse energy to ${\MET}>30$ GeV does
reduce the background of trileptons from heavy flavors by a factor of 10, but
$bZ/\gamma$ is still 10 times the SU2 signal.  Adding the same angular cuts we
suggest in Sec.\ \ref{sec:hffs}, $\theta^\mathrm{CM}_{12}>45^\circ$,
$\theta^\mathrm{CM}_{13}>40^\circ$, and $\theta^\mathrm{CM}_{23}<160^\circ$,
reduces the background of leptons from heavy flavor decays by 30\% with little
effect on the signal.

\begin{table}[ht]
\caption{Leading order signal and background statistics for 30 \fbi\ in an
ATLAS-like analysis, and with two additional cuts.  The virtual photon
components are large after cuts.  For point SU2 we simulate only $\Zb \Wpa$
production.  ATLAS considers $\Zc \Wpa$ production as well, which would
roughly double the SUSY cross section.
\label{tab:ATLres}}
\begin{tabular}{lD{.}{.}{2}D{.}{.}{2}D{.}{.}{2}D{.}{.}{2}}
& \multicolumn{1}{c}{$N^l=3$}& & & \multicolumn{1}{c}{Angular} \\
Channel &\multicolumn{1}{c}{no jets}&
\multicolumn{1}{c}{$M_{ll}^\mathrm{OSSF}<75$~GeV} &
\multicolumn{1}{c}{${\MET > 30}\textrm{ GeV}$} &
\multicolumn{1}{c}{cuts} \\ \hline
SU2 & 38 & 29 & 22 & 21 \\ \hline
$WZ/\gamma$ & 1450 & 387 & 292 & 272 \\
$t\bar t$ & 942 & 466 & 409 & 396 \\
$tW$ & 225 & 116 & 100 & 98 \\
$t\bar b$ & 0.95 & 0.73 & 0.60 & 0.56 \\
$bZ/\gamma$ & 12700 & 3200 & 226 & 138 \\
$cZ/\gamma$ & 3080 & 571 & 36 & 26 \\
$b\bar bZ/\gamma$ & 7760 & 882 & 88 & 75 \\
$c\bar cZ/\gamma$ & 4110 & 745 & 53 & 38 \\
$b\bar bW$ & 8.2 & 5.9 & 4.6 & 4.3 \\
$c\bar cW$ & 0.16 & 0.12 & 0.09 & 0.08
\end{tabular}
\end{table}

The SU2 fixed point region of supersymmetry studied by ATLAS is unlikely to
produce a measurable trilepton signature.  Summing over both $\Zb \Wpa$ and
$\Zc \Wpa$ production could require 450 \fbi\ of data for a $5\sigma$
discovery.  Nevertheless, other regions of SUSY parameter space, such as those
in the CMS study we examine, are viable and share the same backgrounds we
present here.  Our intent is to demonstrate that even under the differently
optimized cuts applied by the ATLAS study, the backgrounds to trileptons from
heavy flavors completely dominate the sample unless additional cuts are made
that explicitly target their removal.  A missing transverse energy cut that
balances the loss of SUSY trilepton signal against background removal, and
cuts based on angular correlations, are both effective means of controlling
the background of leptons from heavy flavor decays.

\section{Conclusions}
\label{sec:conclusions}

In this paper we investigate standard model sources of isolated three lepton
final states at LHC energies.  We provide quantitative estimates of the rates
and kinematic distributions in phase space of events that arise from several
processes.  One of these is the associated production of $WZ$, along with its
generalizations $W\gamma^{*}$, where $\gamma^{*}$ is a virtual photon that
decays as $\gamma^{*}\rightarrow l\bar{l}$.  A major new contribution is the
demonstration that bottom and charm meson decays produce isolated three-lepton
events that can overwhelm the effects of other processes.  We compute
contributions from a wide range of SM heavy flavor processes including $b
Z/{\gamma^{*}}$, $c Z/ \gamma^{*}$, $b \bar{b} Z/ \gamma^{*}$, $c \bar{c} Z/
\gamma^{*}$.  We also include contributions from $t \bar{t}$ production, and
from processes in which a $W$ is produced in association with one or more
heavy flavors such as $t W$, $b \bar{b} W$, $c \bar{c} W$.  In all these
cases, one or more of the final observed isolated leptons comes from a heavy
flavor decay.  These heavy flavor sources dominate the isolated lepton
spectrum at small $p_T$, and these sources of background must be considered in
the evaluation of the significance of any signal that has low-$p_T$ leptons.

Trileptons from $WZ$ production have been examined previously, but we find
that the $W\gamma^{*}$ contribution is much more significant than is widely
appreciated, as has also been noted by others
\cite{backgrounds,Campbell:1999ah}.  Unlike $WZ$, the $W\gamma^{*}$
contribution cannot be reduced by an antiselection of events in which the
invariant mass of the $l\bar{l}$ system is in the vicinity of the $Z^0$ peak.

The dominant behavior of the SM sources motivates the investigation of new
selections (cuts) on the final state kinematic distributions that would be
effective in reducing the backgrounds.  One of these cuts involves selections
on the opening angles among the three charged leptons in the final state.  Our
studies identify specific distributions that can be examined once real data
are available and used to constrain both the shape and magnitude of the
standard model backgrounds.  The strongest of these is the steeply falling
distribution in missing transverse energy.

We use MadEvent~\cite{Maltoni:2002qb} to compute the full matrix elements for
partonic subprocesses that result in a $lll\nu$ final state.  This method
allows us to retain important angular correlations among the final leptons.
These parton level events are then passed through a PYTHIA showering Monte
Carlo code, and finally through a modified PGS detector simulation program.
We compare our overall results with studies done by the CMS and ATLAS groups.

To examine signal discrimination (and to compute signal to background ratios),
we choose as a signal process the supersymmetry example of chargino and
neutralino pair production.  We focus on the SUSY parameter space points LM1,
LM7, and LM9 considered by the CMS Collaboration~\cite{CMSTDR2,CMSpaper} and
on the SU2 point studied by the ATLAS Collaboration~\cite{ATLANAL}.  Some of
these points are expected to have favorable SUSY cross sections at the LHC.

Using the new additional cuts on $\MET$ and angles that we suggest in this
paper, we show that even after the heavy flavor backgrounds are taken into
account, it should be possible to find an approximately $4\sigma$ excess in 30
\fbi\ for SUSY point LM9 and somewhat smaller significances for points LM7 and
LM1.  However, we acknowledge that there are limitations inherent in modeling
of events and simulations of detector response.  These uncertainties are
difficult to evaluate quantitatively.  In particular, we note that our SM
model cross sections are computed at leading order and then passed through the
PYTHIA showering code.  Accompanying hadronic radiation is generated by
PYTHIA, and jet vetoes (no jets with $E_T>30$ GeV) are then applied in the
analysis, removing some fraction of the cross section.  PYTHIA currently
treats events generated internally and events fed into it differently when
producing initial state radiation.  The jet veto is very sensitive to the
details of this radiation spectrum, and hence there is uncertainty in the
overall normalization of the events we simulate.  A dedicated study of these
differences in PYTHIA should be performed before ATLAS and CMS attempt to fit
their measurements of initial state radiation and of underlying events.
Regardless of the normalization, this showering issue has little effect on the
ratio of leptonic events coming from heavy flavor decays to those from
$WZ/\gamma^*$ or $t\bar t$.

An alternative approach would begin with next-to-leading order matrix elements
and a showering code that deals properly with matching and double counting
aspects of the radiation.  Not having this tool available, and recognizing
that it also will have its limitations, we provide an alternative (and more
optimistic) assessment of the signal significance based on a partially {\em ad
hoc} approach.  The CMS study finds that, as a result of the jet veto applied
in the analysis, the trilepton background obtained from a next-to-leading
order computation of the $t \bar{t}$ final state is about a factor of 3.6
smaller than one would obtain from a leading-order computation.  Applying this
factor of 3.6 universally to all the SM backgrounds we compute, we can obtain
an approximately $8\sigma$ excess in 30 \fbi\ for SUSY point LM9.  We will not
know which estimate of initial state hadronic radiation is more correct until
it is measured at the LHC.

The analysis for the ATLAS focus point SU2 is similar to the CMS analysis.
The spectrum of the model near SU2 is exceptionally sensitive to assumptions
regarding the top-quark mass, and it produces too few events to observe.
Nevertheless, the analysis is applicable across a broad range of SUSY
parameters that produce trilepton signatures, and the backgrounds remain the
same. By default ATLAS applies an additional modest missing transverse energy
cut and angular correlation.  However, the conclusion remains that the
backgrounds including leptons from heavy flavor decays are still much larger
than $WZ/\gamma^*$ or $t\bar{t}$.

In general we find that the dominant backgrounds to low-momentum trilepton
signatures come from real $b$ and $c$ decays.  In the CMS and ATLAS
supersymmetric analyses we examine, the $Z/\gamma^*+$heavy flavor decay
backgrounds are a factor of 10--30 larger than $WZ/\gamma^*$ or $t\bar{t}$ to
trileptons.  Large $\MET$ cuts and angular correlations can be used to
significantly reduce the heavy flavor backgrounds, but we must be mindful of
the modest $\MET$ in the SUSY signal.  Coupled with the results for dileptons
in Ref.\ \cite{Sullivan:2006hb}, it is clear that leptons from heavy flavor
decays should be examined for all low-momentum lepton signals.  Once
normalizations are measured \textit{in situ}, we have handles to reduce the
effect of these backgrounds to an acceptable level.

\begin{acknowledgments}
We thank Tim Stelzer for sharing his improved phase-space code.  Z.\ S.\ is
supported by the U.~S.\ Department of Energy under Contract No.\
DE-FG02-04ER41299, and is a visitor at Argonne National Laboratory.  E.\ L.\
B.\ is supported by the U.~S.\ Department of Energy under Contract No.\
DE-AC02-06CH11357 and thanks the Kavli Institute for Theoretical Physics
(KITP), Santa Barbara, for hospitality during the final stages of this work.
The KITP is supported by the National Science Foundation under Grant No.\ NSF
PHY05-51164.  We gratefully acknowledge the use of JAZZ, a 350-node
computer cluster operated by the Mathematics and Computer Science Division at
Argonne as part of the Laboratory Computing Resource Center.
\end{acknowledgments}

\appendix*
\section{Isolated lepton reconstruction}
\label{app:isolep}

The isolation criteria for muons and electrons used in this paper are based on
the ATLAS working definitions that were publicly available at the time of
this study~\cite{ATLLEPS}.  While final criteria will likely change, we
explain in Sec.\ \ref{sec:physbd} and Ref.\ \cite{Sullivan:2006hb} why the
rate of isolated leptons from $b$ and $c$ decay is fairly stable.  We use
these same criteria to define isolated leptons for the CMS study.  We apply
additional cuts on transverse momentum for the CMS and ATLAS studies as
described in Secs.\ \ref{sec:CMS} and \ref{sec:ATLAS}.

A muon is said to be isolated if there is a charged track with $p_{T\mu}>10$
GeV, a hit in the muon chamber at $|\eta_\mu|<2.5$, and two isolation cuts are
passed.  The sum of the transverse momentum of all other tracks in a cone of
size $\Delta R = 0.2$ must be less than an isolation energy of 4 GeV, and the
sum of the energies in all calorimeter towers surrounding the one containing
the muon that lie within a cone of size $\Delta R = 0.4$ must be less than 10
GeV.  Finally, a pseudorapidity-dependent detector efficiency is applied that
averages over cracks, noise, etc., of $\approx 0.836$ for $|\eta_\mu|<1.05$,
and $\approx 0.922$ for $1.05<|\eta_\mu|<2.5$.

Electron reconstruction is based on first defining ``regions of interest''
(ROI).  A fixed $0.1\times 0.1$ window in $\eta$--$\phi$ space is scanned over
electromagnetic calorimeter towers in the region $|\eta|<2.5$ and
$0\le\phi<2\pi$ and used to identify ROI with $E_T^{\mathrm{EM}}>10$ GeV and
$E_T^{\mathrm{had}}<2$ GeV.  The segmentation of the ATLAS electromagnetic
(EM) calorimeter leads to towers of approximately $0.05\times 0.05$.  The
energy in the twelve towers surrounding the four towers of the ROI are added,
and isolation requires $E_{12}<3$ GeV.  Finally, we require the ROI to have a
track within a cone of $0.1$ with $0.7 < E_T^{\mathrm{EM}}/p_T^{\mathrm{trk}}
< 1.4$, and apply an overall efficiency of $\approx 0.723$.

\end{document}